\documentclass{iau} 
\usepackage{graphicx}

\title[Mass transfer in  asymptotic-giant-branch binary systems] 
{Mass transfer in  asymptotic-giant-branch binary systems}

\author[Z. Chen, A. Frank, E. G. Blackman, J. Nordhaus, J. Carroll-Nellenback]   
{Zhuo Chen$^1$, Adam Frank$^1$, Eric G. Blackman$^1$, Jason Nordhaus$^2$, \and Jonathan Carroll-Nellenback$^1$
}

\affiliation{$^1$Department of Physics and Astronomy, University of Rochester, Rochester, NY 14627\\ email: {\tt zchen25@ur.rochester.edu} \\[\affilskip]
$^2$National Technical Institute for the Deaf, Rochester Institute of Technology, Rochester, NY 14623}

\pubyear{2016}
\volume{323}  
\jname{Planetary Nebulae}
\editors{X. Liu, L. Stanghellini, \& A. Karakas. Editor, eds.}
\begin{document}

\maketitle

\begin{abstract}
Binary stars can interact via mass transfer when one member (the primary) ascends onto a giant branch. The amount of gas ejected by the binary  and the amount of gas accreted by the secondary over the lifetime of the primary influence the subsequent binary phenomenology. Some of the gas ejected by the binary will remain gravitationally bound and its distribution will be closely related to the formation of planetary nebulae. We investigate the nature of mass transfer in binary systems containing an AGB star by adding radiative transfer to the \emph{AstroBEAR AMR Hydro/MHD} code.
\end{abstract}

{\bf Introduction:} Binaries are believed to be the progenitors of Type Ia supernovae, planetary nebulae and many other systems \cite[(Nordhaus \& Blackman 2006; Ivanova 2013)]{nordhaus2006,Ivanova2013}. In addition, many post-AGB stars show evidence for disks which must due to mass transfer with a companion. Our primary focus in this work is to study PNe projenitor binary systems with different orbital radii to explore changes in mass transfer. 

\begin{figure}[!ht]
    \centering
    \includegraphics[width=6.6cm]{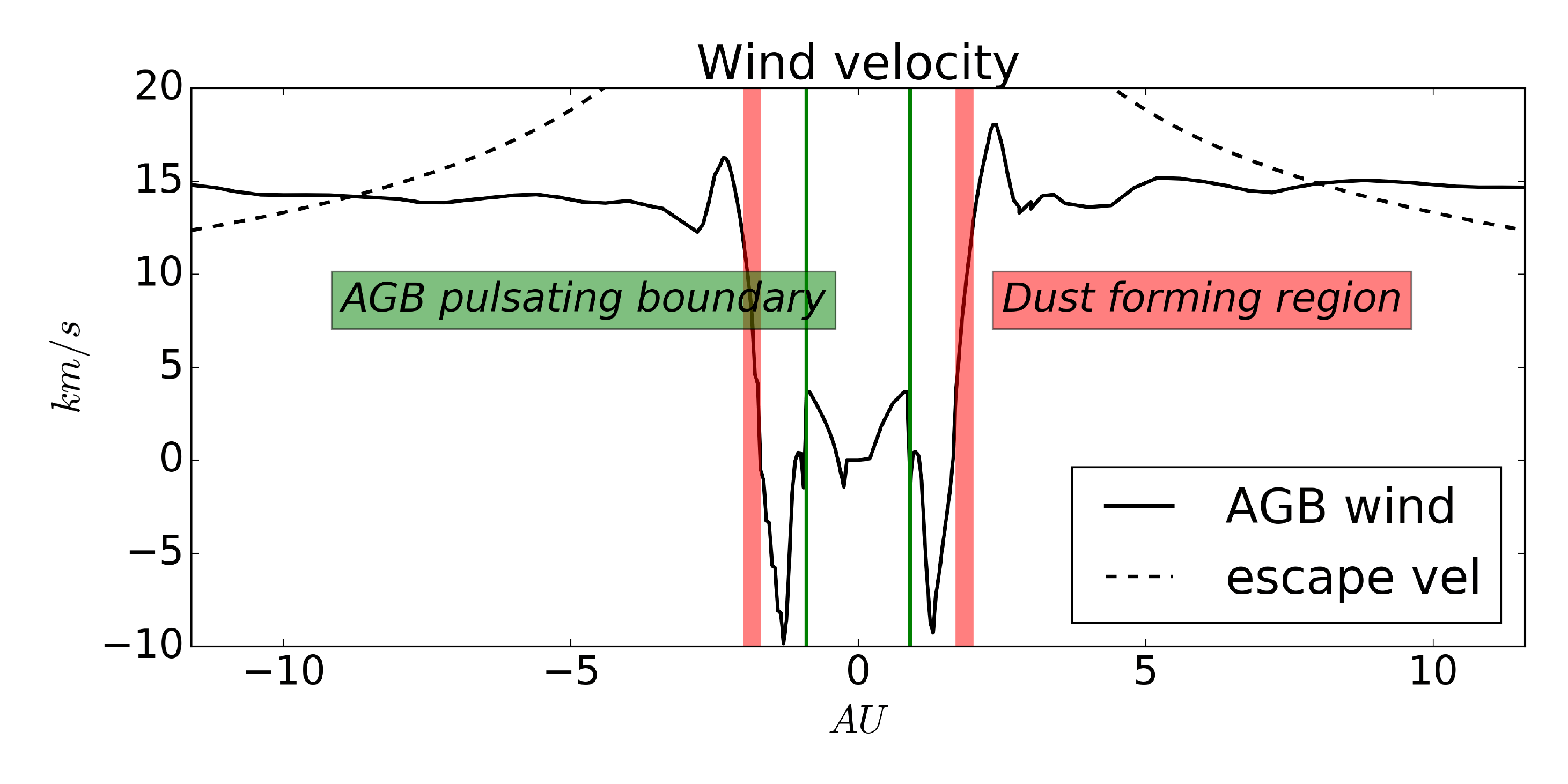}
    \includegraphics[width=6.6cm]{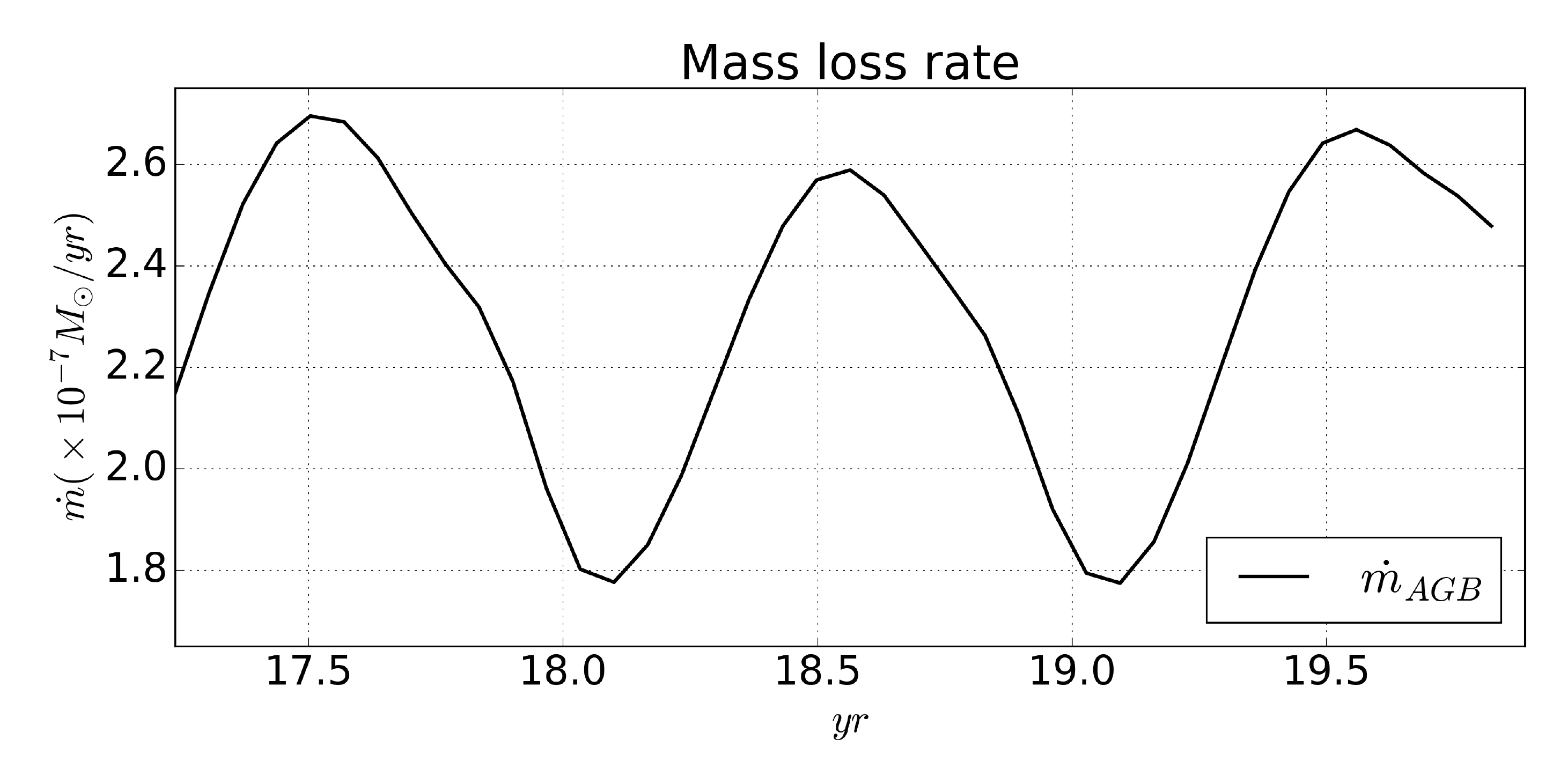}
    \caption{AGB wind's velocity profile and mass-loss rate.}
    \label{singleAGB}
\end{figure}

{\bf Model:}We first test our AGB-wind mass-loss model using an isolated AGB star. We model the AGB star with a pulsating ($P_{pulse}=365d$), luminous boundary and point gravity
\cite[(Bowen 1988)]{bowen1988}. We assume dust forms when the equilibrium temperature is below the condensation temperature. The dust will change the opacity of the gas.  We show in Fig. \ref{singleAGB} that our AGB wind model produces a physically reasonable terminal velocity of $15\ km/s$ and  average mass loss of $2.3\times10^{-7}\ M_{\odot}/yr$.

\begin{figure}[!ht]
\begin{center}
 \includegraphics[width=3.2cm]{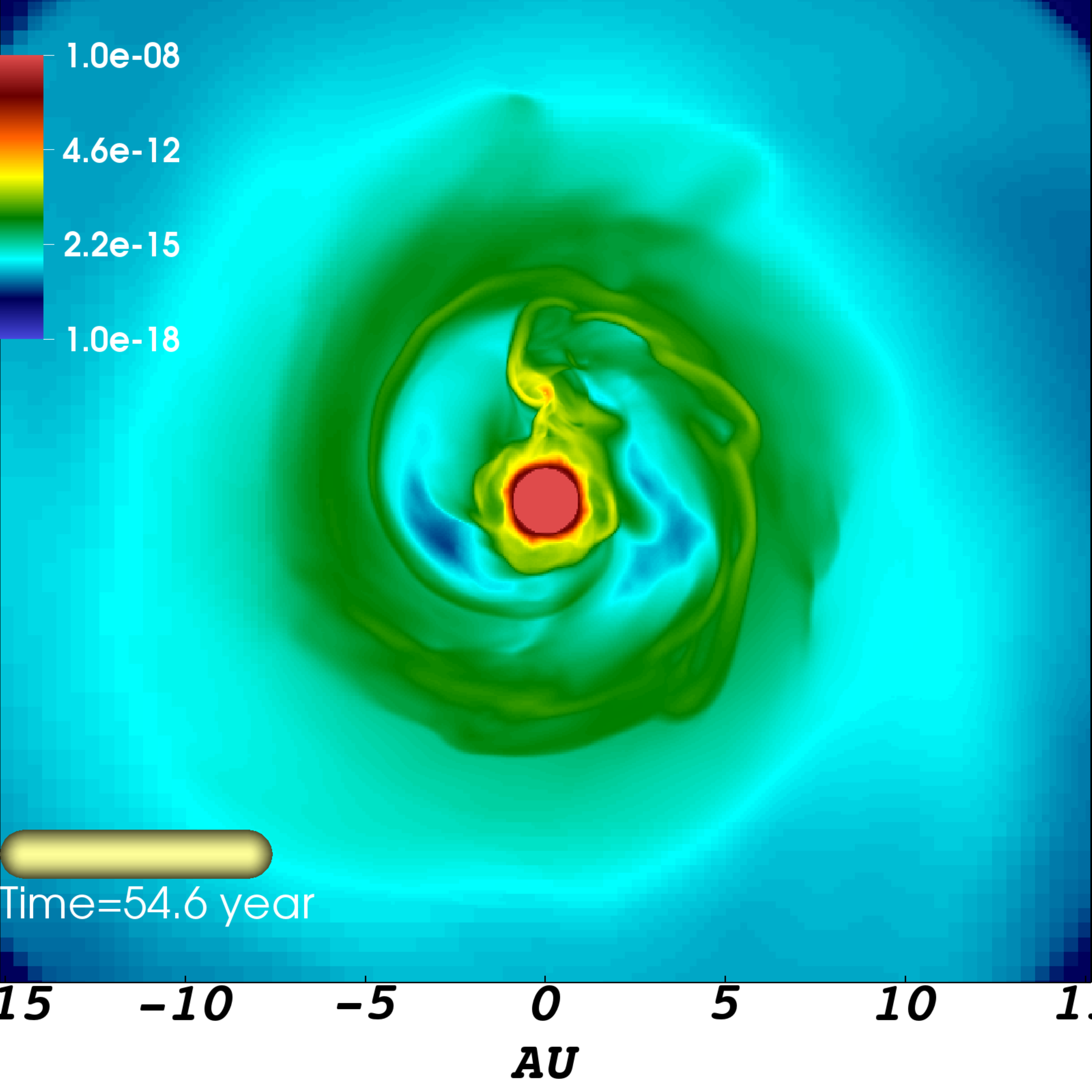}
 \includegraphics[width=3.2cm]{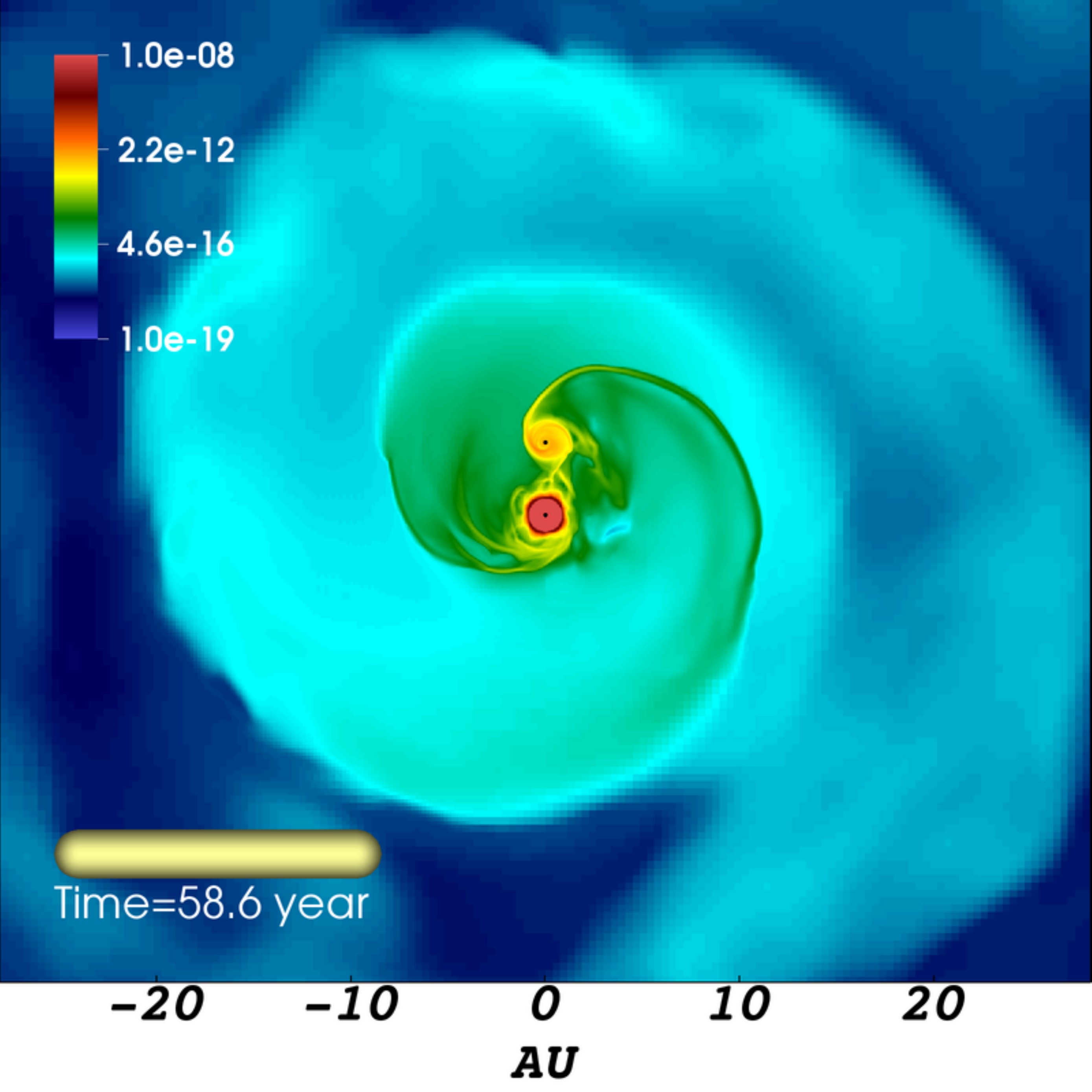}
 \includegraphics[width=3.2cm]{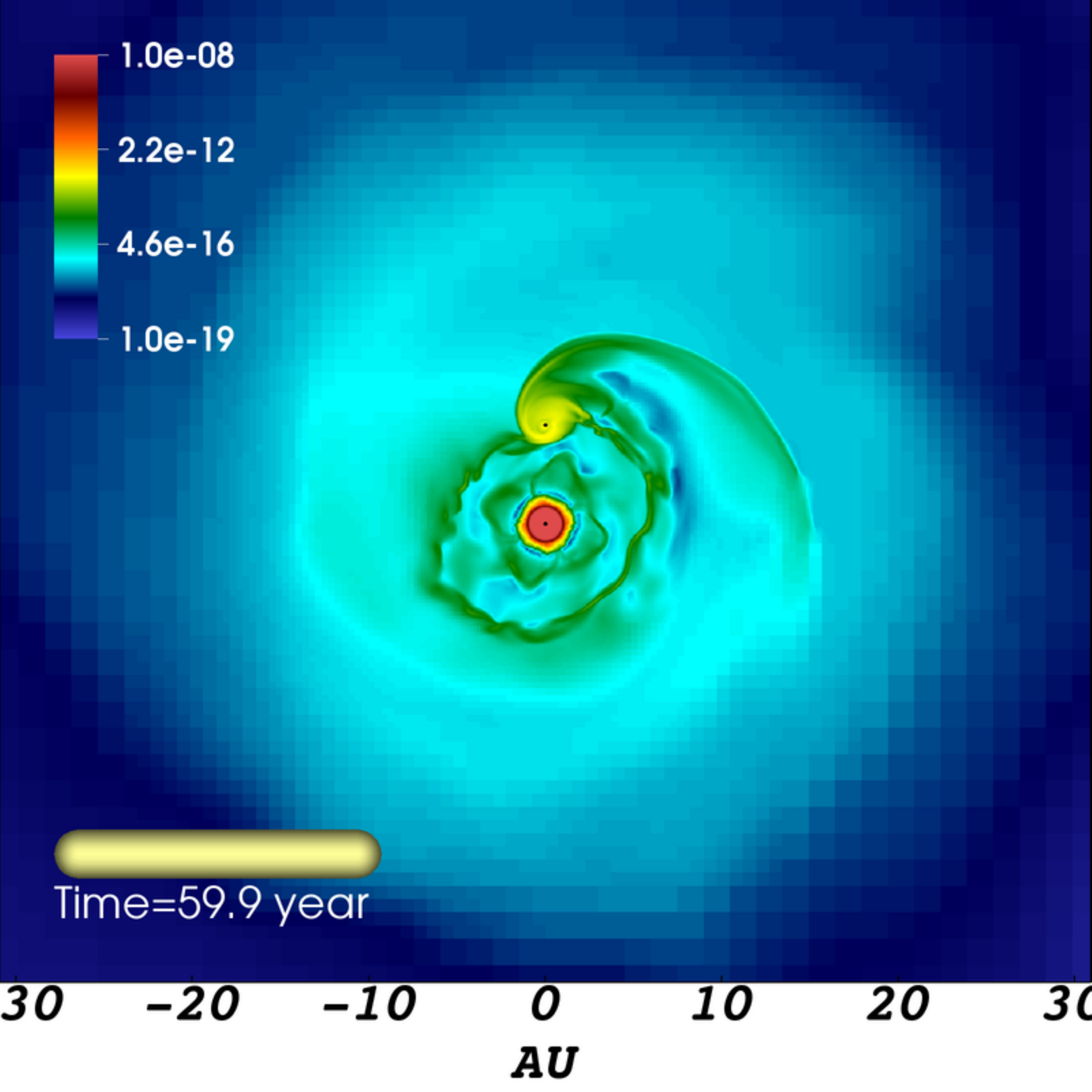}
 \includegraphics[width=3.2cm]{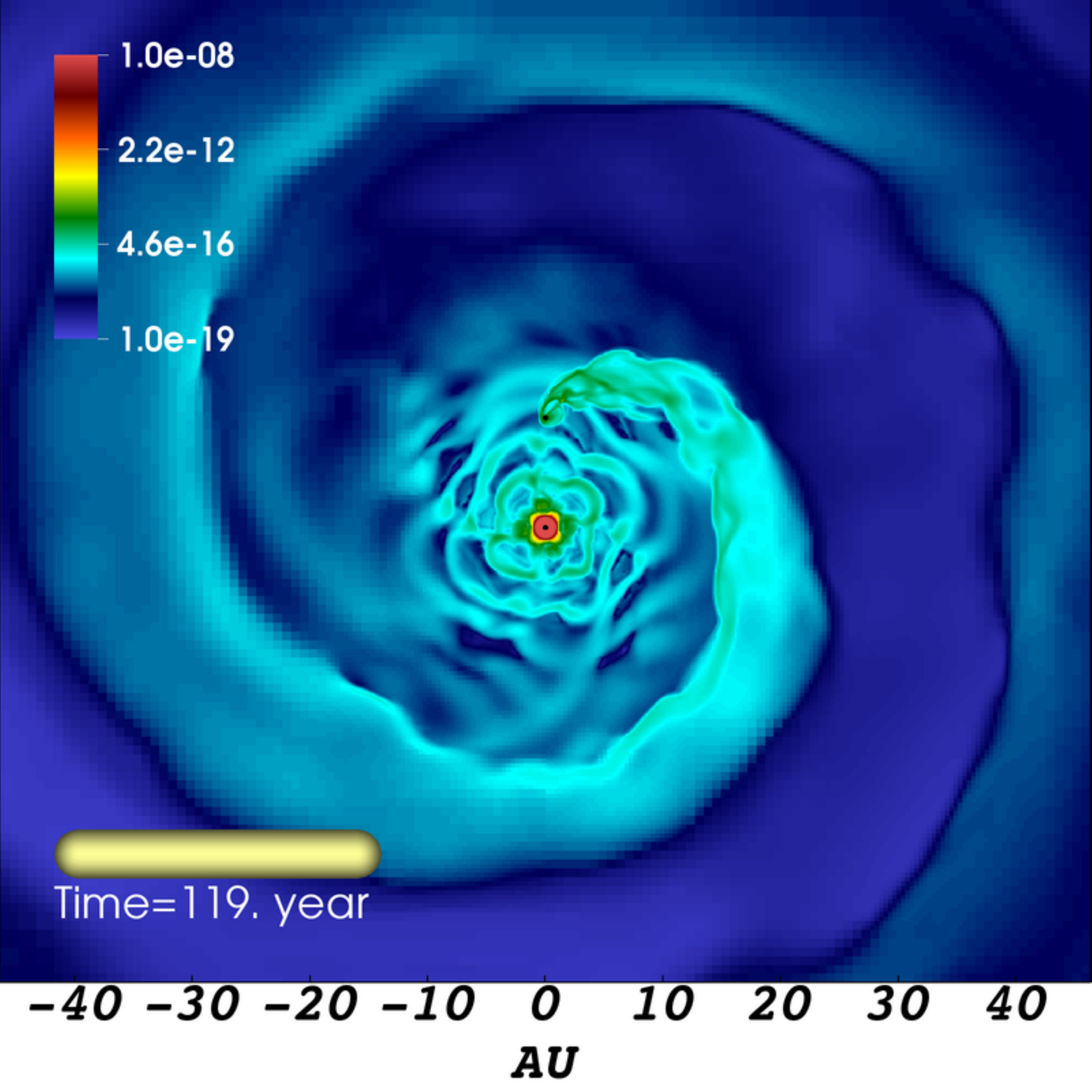}\\
 \includegraphics[width=3.2cm]{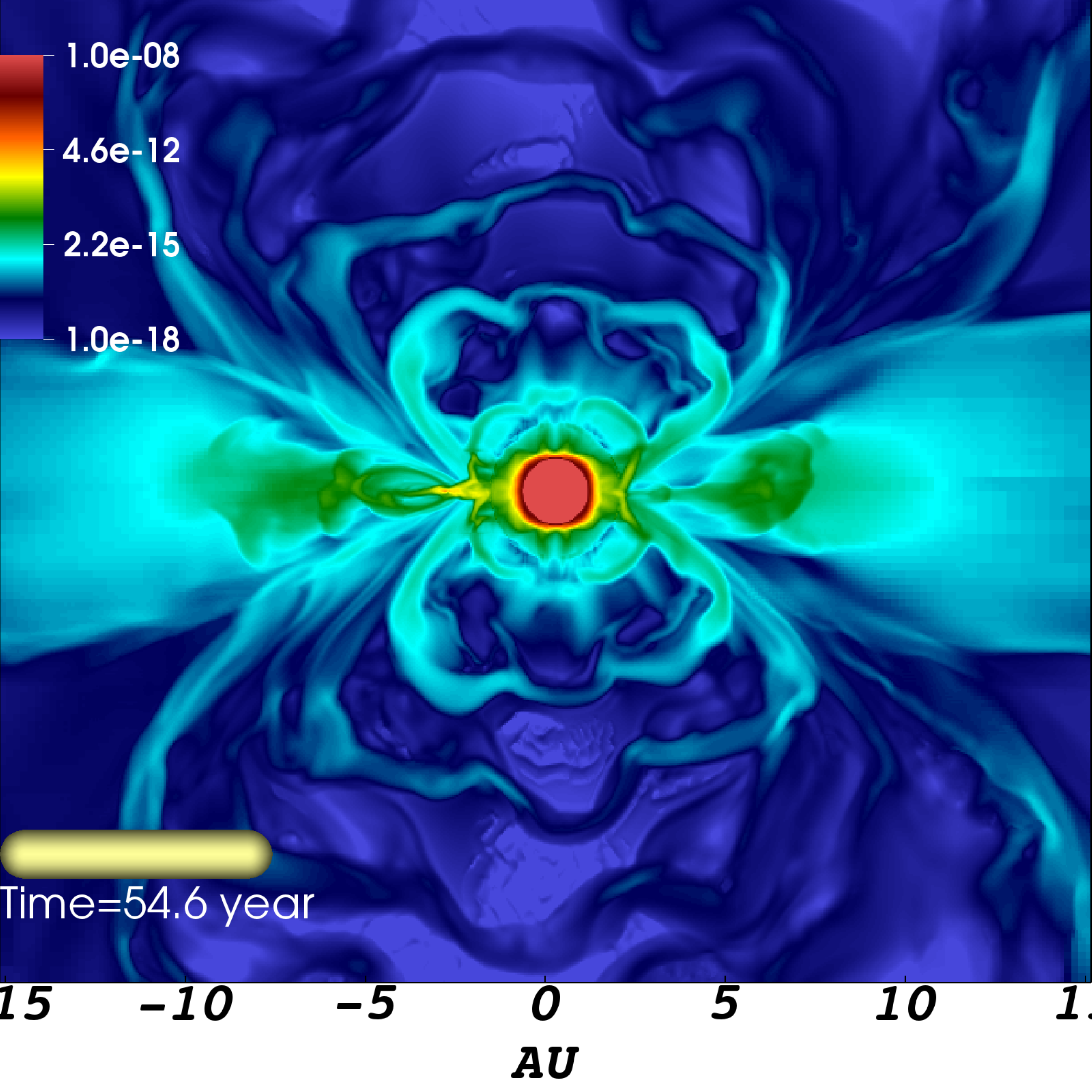}
 \includegraphics[width=3.2cm]{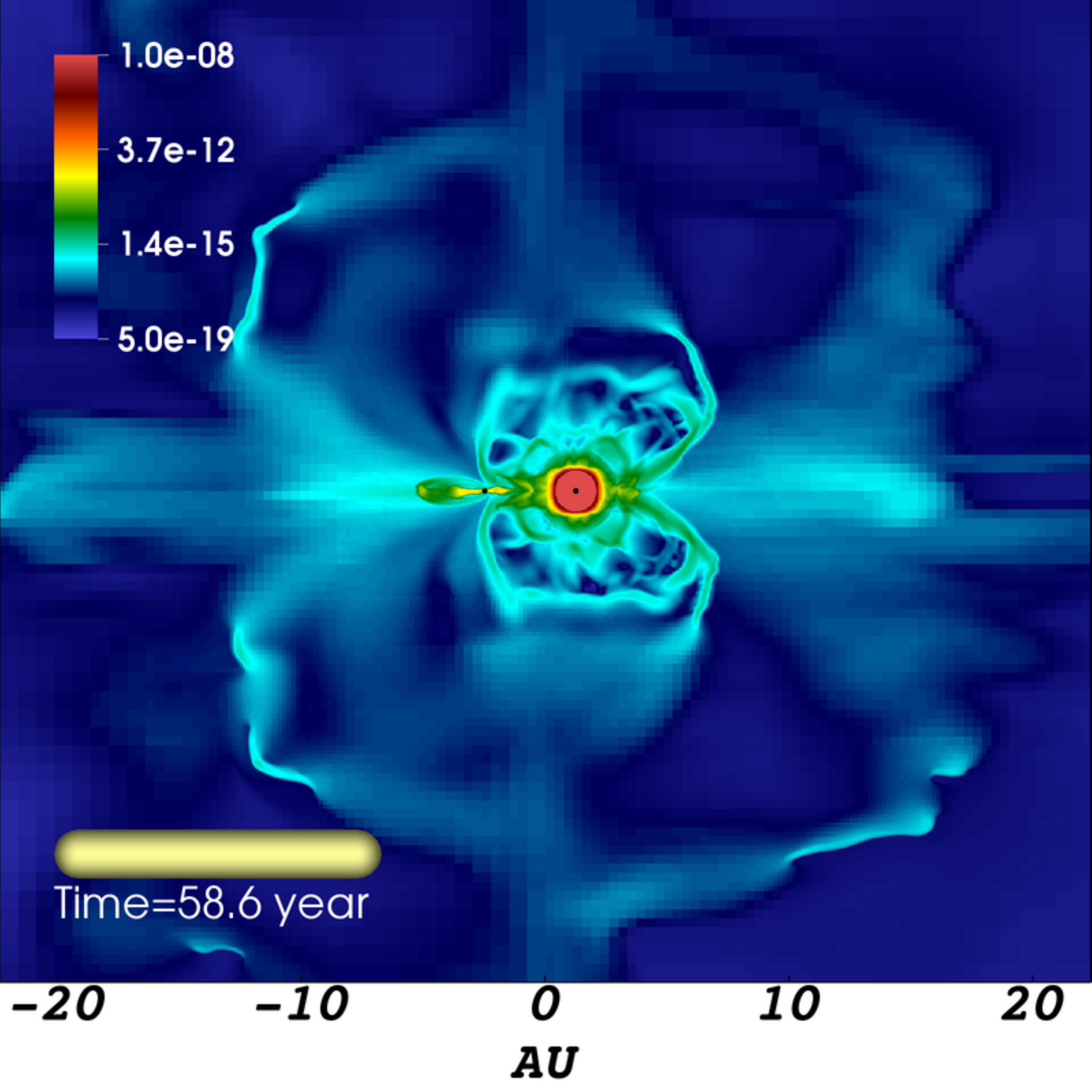}
 \includegraphics[width=3.2cm]{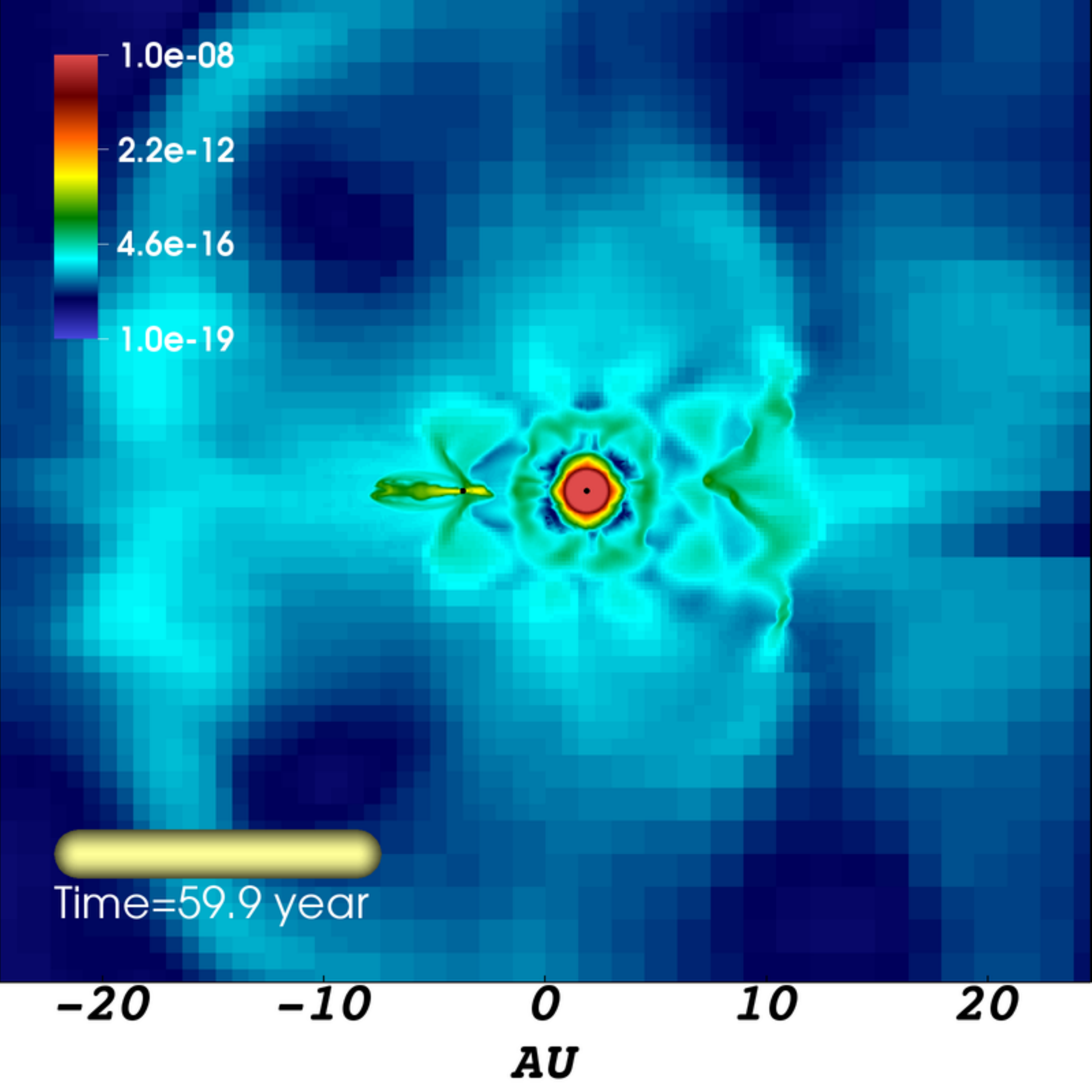} 
 \includegraphics[width=3.2cm]{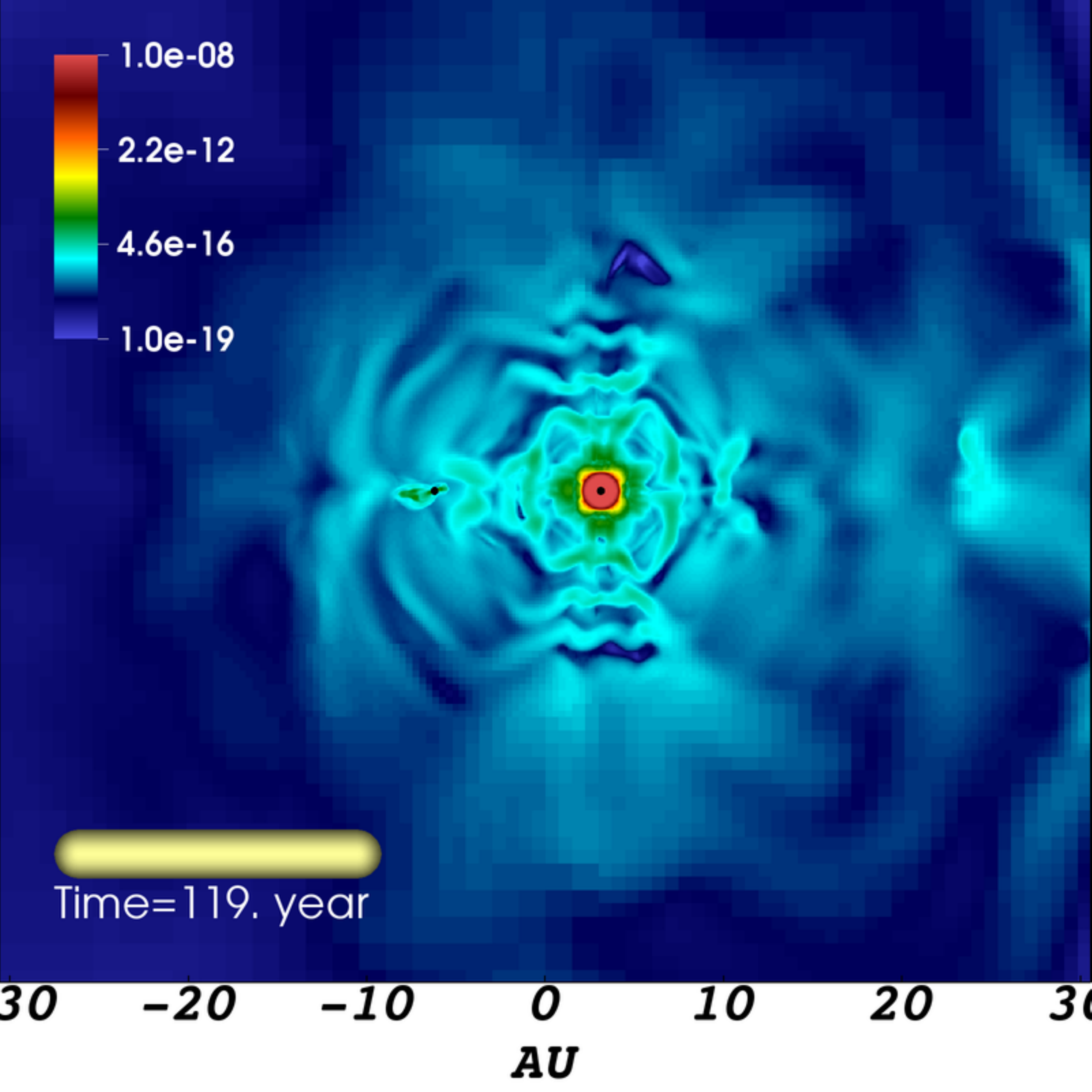}\\
 \caption{From left to right, the figures correspond to each of the four models listed in table \ref{masslosstable}. Figures on top show the color density plot in $g/cm^3$ of the middle z-plane cut. Figures on bottom show the middle x-plane cut.}
 \label{binary}
\end{center}
\end{figure}

We then place an accreting secondary
\cite[(Krumholz et al. 2004)]{krumholz2004} in circular orbit with the AGB star. The boundary condition of the AGB star is kept the same as in the isolated case discussed above. However,  the mass of the secondary and the separation of the binary are allowed to vary. We calculate the affected mass-loss rate of the AGB star, the accretion rate of the secondary (Table \ref{masslosstable}), and examine the morphology of the outflows (Fig. \ref{binary}).

\begin{table}[!ht]
    \centering
    \begin{tabular}{|c|c|c|c|c|c|}
    \hline
        model\# & $m_{AGB}(M_{\odot})$ & $m_{sec}(M_{\odot})$ & $d(AU)$ & $\dot{m}_{AGB}\left(M_{\odot}/yr\right)$ & $\dot{m}_{sec}\left(M_{\odot}/yr\right)$ \\ \hline
    1 & 1.0 & 0.1 & 3  & $3.4\times10^{-7}$ & $-1.3\times10^{-7}$ \\ \hline
    2 & 1.0 & 0.5 & 4  & $3.1\times10^{-7}$ & $-1.0\times10^{-7}$ \\ \hline
    3 & 1.0 & 0.5 & 6  & $2.8\times10^{-7}$ & $-4.8\times10^{-8}$ \\ \hline
    4 & 1.0 & 0.5 & 10 & $2.6\times10^{-7}$ & $-5.8\times10^{-9}$ \\
    \hline
    \end{tabular}
    \caption{The affected mass loss rate of the AGB star and the secondary.}
    \label{masslosstable}
\end{table}
\vspace{-0.2cm}

{\bf Results:}
1. The AGB mass-loss rate is changed by the presence of the accreting secondary.

2. The accretion rate onto the secondary can increase by up to $\sim$38\% of the mass-loss rate. This exemplifies Wind-Roche-lobe-overflow
\cite[(Podsiadlowski \& Mohamed 2007)]{podsiadlowski2007}.

3. Circumbinary disks and accretion disks (around the secondary) are conspicuously found in our calculations.

4. The resulting aspherical circumbinary density distribution provides insight for understanding aspects of  aspherical planetary nebulae \cite[(Maercker et al. 2012)]{maercker2012}.


\vspace{-0.3cm}
\begin{thebibliography}{}

\bibitem[Bowen (1988)]{bowen1988}{Bowen, G. H.} 1988, \textit{ApJ}, 329, 299

\bibitem[Ivanova et al. (2013)]{Ivanova2013}{Ivanova N, J.ustham S., Chen X., De Marco O., Fryer C.L., Gaburov E., Ge H., Glebbeek E., Han Z., Li X.D., \& Lu G.} 2013, \textit{A\&AR}, 21(1), 1-73

\bibitem[Krumholz et al. (2004)]{krumholz2004}{Krumholz, Mark R., Christopher F. McKee, \& Richard I. Klein.} 2004, \textit{ApJ}, 611, 399

\bibitem[Maercker et al. (2012)]{maercker2012}{Maercker M., Mohamed S., Vlemmings W.H., Ramstedt S., Groenewegen M.A., Humphreys E., Kerschbaum F., Lindqvist M., Olofsson H., Paladini C., \& Wittkowski M.} 2012, \textit{Nature}, 490, 232

\bibitem[Nordhaus \& Blackman(2006)]{nordhaus2006} Nordhaus, J., \& Blackman, E.~G.\ 2006, \textit{MNRAS}, 370, 2004 

\bibitem[Podsiadlowski \& Mohamed (2007)]{podsiadlowski2007}{Podsiadlowski, Ph., \& S. Mohamed.} 2007, \textit{Baltic Astronomy}, 16, 26


\end{thebibliography}
\end{document}